# Mechanisms behind the Generalized Synchronization Conditions

A. A. Koronovskii, O. I. Moskalenko, and A. E. Hramov

*Chernyshevsky State University, Saratov, 410012 Russia*
*e-mail: alkor@cas.ssu.runnet.ru; aeh@cas.ssu.runnet.ru*

**Abstract**—A universal mechanism underlying generalized synchronization conditions in unidirectionally coupled stochastic oscillators is considered. The consideration is carried out in the framework of a modified system with additional dissipation. The approach developed is illustrated with model examples. The conclusion is reached that two types of the behavior of nonlinear dynamic systems known as generalized synchronization and noise-induced synchronization, which are viewed as different phenomena, actually represent a unique type of the synchronous behavior of stochastic oscillators and are caused by the same mechanism.

## INTRODUCTION

Synchronization of oscillations, a most important nonlinear phenomenon, which is attracting considerable attraction [1], is of both scientific and applied interest (e.g., in biology and physiology [2], secure data transmission by means of chaotic signals [3–5], control of microwave electronic devices [6], etc.).

The development of the theory of dynamic chaos has made it possible to reveal several types of the chaotic synchronous behavior of dynamic systems [7, 8]. Among them are phase synchronization [7], generalized synchronization [9], lag synchronization [10], intermittent lag synchronization [11], intermittent generalized synchronization [12], and complete synchronization [13], each having its own features and types of diagnostics. The interplay between these types of synchronization is the subject of wide speculation. Various types of synchronization between randomly coupled oscillators can be viewed as various manifestations of a unified law which coupled nonlinear systems obey (see, e.g., [14–18]). In [16, 17, 19], a new type of the synchronous behavior of stochastic oscillators has been brought into consideration, time-scale synchronization, which in a natural way generalizes the synchronizations listed above.

Generalized synchronization of unidirectionally coupled oscillators, a sort of the synchronous chaotic behavior, has been attracting special attention [9]. Here, the states of unidirectionally coupled interacting driving, $\mathbf{x}_d(t)$, and driven, $\mathbf{x}_{dr}(t)$, stochastic oscillators (with discrete or continuous time) are related by some function $\mathbf{F}[\cdot]$ such that the relation $\mathbf{x}_{dr}(t) = \mathbf{F}[\mathbf{x}_d(t)]$ is established after the transient has been completed.

The form of function $\mathbf{F}[\cdot]$ (smooth or fractal) may be rather complicated, and the procedure of its finding may be nontrivial. Strong and weak generalized synchronizations are distinguished [20]. It should be noted that two different dynamic systems, including those of different phase space dimensions, may serve as interacting oscillators.

Several methods of diagnosing generalized synchronization between stochastic oscillators, such as the nearest neighbors method [9, 21] and the frequently used method of auxiliary system [22], have been suggested. The essence of the latter is the following: along with driven system $\mathbf{x}_{dr}(t)$, auxiliary system $\mathbf{x}_a(t)$ identical to it is introduced. Initial conditions for auxiliary system $\mathbf{x}_a(t_0)$ are taken other than those for driven system $\mathbf{x}_{dr}(t_0)$ but lying in the range of attraction of the same attractor. If generalized synchronization between the interacting systems is absent, vectors $\mathbf{x}_{dr}(t)$ and $\mathbf{x}_a(t)$ of states of the driven and auxiliary systems belong to the same stochastic attractor but are different. If, however, generalized synchronization takes place, the states of the driven and auxiliary systems must be identical ($\mathbf{x}_{dr}(t) \equiv \mathbf{x}_a(t)$) (after the transient has been complete) by virtue of the relationships $\mathbf{x}_{dr}(t) = \mathbf{F}[\mathbf{x}_d(t)]$ and, accordingly, $\mathbf{x}_a(t) = \mathbf{F}[\mathbf{x}_d(t)]$ (for details, see [22]). Thus, the equivalence of the states of the driven and auxiliary systems after the transient (which may be fairly long [12]) is a sign of generalized synchronization between driving and driven oscillators.

Analysis of the generalized synchronization conditions involves calculation of conditional Lyapunov exponents [23, 24]. Lyapunov exponents are calculated for the driven system, and, since its behavior is related to the behavior of the driving system, the exponents will differ from those for an autonomous (uncoupled) driven system (therefore, they are called conditional). The sign of generalized synchronization in unidirectionally coupled dynamic systems is the negativity of the major conditional Lyapunov exponent [20, 23]. It should also be noted that the conditions of complete synchronization and lag synchronization in unidirectionally coupled stochastic oscillators are also the partial cases of generalized synchronization [20].

The goal of this work is to reveal and illustrate a universal mechanism behind generalized synchronization in unidirectionally coupled stochastic oscillators. In Sect. 1, we consider an approach used to explain generalized synchronization that is based on introducing a modified system. In Sects. 2 and 3, the occurrence of generalized synchronization in a number of variously coupled model systems is reported. Finally, Sect. 4 is devoted to the interplay between generalized chaotic synchronization and noise-induced synchronization. It is concluded that these two types of synchronization are manifestations of the unique synchronous behavior of stochastic oscillators and occur by the same mechanism.

## 1. MODIFIED SYSTEM METHOD AS APPLIED TO GENERALIZED CHAOTIC SYNCHRONIZATION

Consider the behavior of two unidirectionally coupled stochastic oscillators

$$\begin{aligned}\dot{\mathbf{x}}_d &= \mathbf{H}(\mathbf{x}_d, \mathbf{g}_d), \\ \dot{\mathbf{x}}_{dr} &= \mathbf{G}(\mathbf{x}_{dr}, \mathbf{g}_{dr}) + \varepsilon\mathbf{P}(\mathbf{x}_d, \mathbf{x}_{dr}),\end{aligned} \quad (1)$$

where $\mathbf{x}_{d,\,dr}$ are the vectors of states of the driving and driven systems, respectively; $\mathbf{H}$ and $\mathbf{G}$ specify the vector field of the systems considered; $\mathbf{g}_d$ and $\mathbf{g}_{dr}$ are the vectors of parameters; term $\mathbf{P}$ is responsible for the unidirectional coupling between the systems; and $\varepsilon$ is the strength of coupling between the systems.

If the dimensions of the driving and driven systems are, respectively, $N_d$ and $N_{dr}$, the behavior of the unidirectionally coupled stochastic oscillators given by (1) can be described by a spectrum of Lyapunov exponents $\lambda_1 \geq \lambda_2 \geq \ldots \geq \lambda_{N_d + N_{dr}}$. Since the behavior of the driving system is independent of the state of the driven oscillator, the spectrum of Lyapunov exponents can be divided into two parts: Lyapunov exponents of the driving system, $\lambda_1^d \geq \ldots \geq \lambda_{N_d}^d$, and conditional Lyapunov exponents of the driven system, $\lambda_1^{dr} \geq \ldots \geq \lambda_{N_{dr}}^{dr}$. As was noted above, the sign of generalized synchronization is the change of the minus sign to the plus sign in major exponent $\lambda_1^{dr}$.

As a rule, the generalized synchronization conditions are considered for two identical stochastic oscillators with unidirectional dissipative coupling and slightly differing parameters. We therefore will start with this case. Another type of coupling and also the situations where generalized synchronization takes place for different dynamic systems (in particular, for unidirectionally coupled Ressler and Lorentz systems) will be considered below. For two identical systems with unidirectional dissipative coupling, the dimensions of the phase spaces of the driving and driven systems will be the same ($N_d = N_{dr} = N$) and Eqs. (1) can be recast as

$$\begin{aligned}\dot{\mathbf{x}}_d &= \mathbf{H}(\mathbf{x}_d, \mathbf{g}_d), \\ \dot{\mathbf{x}}_{dr} &= \mathbf{H}(\mathbf{x}_{dr}, \mathbf{g}_{dr}) + \varepsilon\mathbf{A}(\mathbf{x}_d - \mathbf{x}_{dr}),\end{aligned} \quad (2)$$

where $\mathbf{A} = \{\delta_{ij}\}$ is the coupling matrix, $\varepsilon$ is a scalar parameter characterizing the strength of the coupling, $\delta_{ii} = 0$ or 1, and $\delta_{ij} = 0$ if $i \neq j$.

Let us consider driven system $\mathbf{x}_{dr}(t)$ as a nonautonomous modified system,

$$\dot{\mathbf{x}}_m = \mathbf{H}'(\mathbf{x}_m, \mathbf{g}_{dr}, \varepsilon), \quad (3)$$

subjected to external action $\varepsilon\mathbf{A}\mathbf{x}(t)$,

$$\dot{\mathbf{x}}_m = \mathbf{H}'(\mathbf{x}_m, \mathbf{g}_{dr}, \varepsilon) + \varepsilon\mathbf{A}\mathbf{x}_d, \quad (4)$$

where

$$\mathbf{H}'(\mathbf{x}, \mathbf{g}) = \mathbf{H}(\mathbf{x}, \mathbf{g}) - \varepsilon\mathbf{A}\mathbf{x}. \quad (5)$$

Note that the term $-\varepsilon\mathbf{A}\mathbf{x}$ introduces additional dissipation into modified system (3).

Clearly, generalized synchronization occurring in system (2) as coupling parameter $\varepsilon$ grows can be viewed as a consequence of two interrelated concurrent processes: an increase in the dissipation in modified system (3) and an increase in the external signal amplitude. It is obvious that both are related to each other through parameter $\varepsilon$ and cannot be realized in driven system (2) independently of each other. Yet, to gain a deeper insight into the mechanism behind the establishment of generalized synchronization, we will take these processes as independent ones and begin with the autonomous behavior of modified system (3).

For this modified system, parameter $\varepsilon$ serves as a dissipation parameter. If $\varepsilon = 0$, the behavior of modified system $\mathbf{x}_m(t)$ is totally coincident with that of driven system $\mathbf{x}_{dr}(t)$ in the absence of coupling. As $\varepsilon$ grows, the dynamics of modified system (3) is bound to simplify. Accordingly, stochastic oscillations in modified system $\mathbf{x}_m(t)$ are expected to change to regular (periodic) oscillations and the system may even turn into the steady state (when the dissipation parameter is high). In this case, one of the Lyapunov exponents of the modified system, $\lambda_0^m$ equals zero (or is negative if modified system (3) is in the steady state) and the rest of the expo-nents are negative ($0 > \lambda_1^m \geq \ldots \geq \lambda_{N-1}^m$). Note, however, that the spectrum of Lyapunov exponents for system (3) differs from the spectrum of conditional Lyapunov exponents $\lambda_1^{dr} \geq \ldots \geq \lambda_N^{dr}$ for driven system (2). The reason is that, unlike for the modified system, the spectrum of conditional Lyapunov exponents depends on the behavior of both the driven and driving systems (see (2)). Therefore, considering only Lyapunov exponents for the modified system, one cannot reach a conclusion that generalized synchronization is set in initial system (2) of unidirectionally coupled stochastic oscillators.

The external signal in (4), conversely, tends to obtrude the chaotic dynamics of driving system $\mathbf{x}_d(t)$ on modified system $\mathbf{x}_m(t)$ and, thereby, complicate the dynamics of the latter. Clearly, generalized synchronization may exist only if the intrinsic chaotic dynamics of modified system $\mathbf{x}_m(t)$ is suppressed by increasing the dissipation. It is also clear that, only when this condition is fulfilled will the current state of modified system $\mathbf{x}_m(t)$ depend on the external signal, i.e., only then will the relationship $\mathbf{x}_m(t) = \mathbf{F}[\mathbf{x}_d(t)]$ be satisfied. According to (4), the functional relationship $\mathbf{x}_{dr}(t) = \mathbf{F}[\mathbf{x}_d(t)]$ will also hold, which corresponds to the generalized synchronization conditions.

Thus, generalized synchronization in system (2) may take place when parameter $\varepsilon$ is such that modified system (3) demonstrates periodic oscillations or passes into the steady state. At the same time, it is well known that even a periodic external perturbation may generate chaotic dynamics in a system exhibiting periodic behavior. Therefore, the regular steady regime established must be sufficiently stable for an external perturbation not to generate chaotic dynamics in modified system $\mathbf{x}_m(t)$. In other words, the difference between coupling parameters $\varepsilon_{gs}$ (at the time generalized synchronization sets in) and $\varepsilon = \varepsilon_p$ (at the time the modified system changes to periodic oscillations) must be sufficiently large.

Under the conditions of generalized synchronization ($\varepsilon > \varepsilon_{gs}$), the amplitude of the external perturbation turns out to be much smaller than the amplitude of periodic oscillations in modified system $\mathbf{x}_m(t)$ (provided that the oscillations are regular). Then, generalized synchronization in this case may be treated as a weak chaotic external perturbation of the periodic dynamics.

The same conclusion can be drawn when the steady state is established in modified system $\mathbf{x}_m(t)$ at relatively high values of parameter $\varepsilon$. In this case, generalized synchronization is, in essence, a chaotic perturbation of the steady state. In other words, the behavior of the system is a transient process that tends to the steady state disturbed by a chaotic external action. If control parameters $\mathbf{g}_{d, dr}$ of the driving and driven systems differ insignificantly and parameter $\varepsilon$ is sufficiently large, the transient will take a short time and, correspondingly, the representative point in the phase space of modified system $\mathbf{x}_m(t)$ will follow the disturbed "steady state" with small lag $\tau$; that is, the condition of lag synchronization sets in.

## 2. GENERALIZED SYNCHRONIZATION IN CHAOTIC SYSTEMS WITH DISSIPATIVE COUPLING

Let us visualize the approach stated in Sect. 1 by examples of generalized synchronization. Consider first two unidirectionally coupled Ressler oscillators with slightly differing parameters,

$$\begin{aligned}
\dot{x}_d &= -\omega_d y_d - z_d, \\
\dot{y}_d &= \omega_d x_d + a y_d, \\
\dot{z}_d &= p + z_d(x_d - c), \\
\dot{x}_{dr} &= -\omega_{dr} y_{dr} - z_{dr} + \varepsilon(x_d - x_{dr}), \\
\dot{y}_{dr} &= \omega_{dr} x_{dr} + a y_{dr}, \\
\dot{z}_{dr} &= p + z_{dr}(x_{dr} - c).
\end{aligned} \qquad (6)$$

In system (6), parameter $\varepsilon$ characterizes the strength of coupling between the oscillators. The control parameters were the same as in [25]: $a = 0.15$, $p = 0.2$, $c = 10.0$, and $\omega_{dr} = 0.95$.

The modified Ressler system has the form

$$\begin{aligned}
\dot{x}_m &= -\omega_{dr} y_m - z_m - \varepsilon x_m, \\
\dot{y}_m &= \omega_{dr} x_m + a y_m, \\
\dot{z}_m &= p + z_m(x_m - c).
\end{aligned} \qquad (7)$$

Figure 1a shows the bifurcation diagram for modified Ressler system (7). It is seen that, as the dissipation parameter grows, the chaotic oscillations in system (7) change to periodic oscillations (starting at $\varepsilon = \varepsilon_p \approx 0.06$; the arrow in Fig. 1a) through a cascade of period doubling reverse bifurcations.

Figure 1b illustrates the $\varepsilon$ dependence of four major Lyapunov exponents for system (6) of two unidirectionally coupled Ressler oscillators with a small offset of control parameter $\omega$ ($\omega_d = 0.99$). Two of them, $\lambda_1^d$ and $\lambda_2^d$, govern the behavior of the driving oscillator and are therefore $\varepsilon$ independent. Two others, $\lambda_1^{dr}$ and $\lambda_2^{dr}$, characterize the behavior of the driven system, depend on coupling parameter $\varepsilon$, and are Lyapunov conditional exponents. If $\varepsilon = 0$ in this case, exponents $\lambda_1^{dr}$ and $\lambda_2^{dr}$ coincide with the Lyapunov exponents of the modified system, $\lambda_1^m$ and $\lambda_2^m$. Since the modified system exhibits chaotic oscillations at $\varepsilon = 0$ (Fig. 1a), the major parameter among the two conditional exponents is positive ($\lambda_1^{dr} > 0$) and the other equals zero ($\lambda_2^{dr} = 0$). As $\varepsilon$ grows, the second conditional Lyapunov exponent becomes negative ($\varepsilon \approx 0.04$) but the dynamics of the modified system remains chaotic, as indicated by the positiveness of major conditional Lyapunov exponent $\lambda_1^{dr}$. As $\varepsilon$ increases further ($\varepsilon_p \approx 0.06$), the behavior of the modified system becomes periodic (Fig. 1a) but generalized synchronization in system (6) is not yet established.

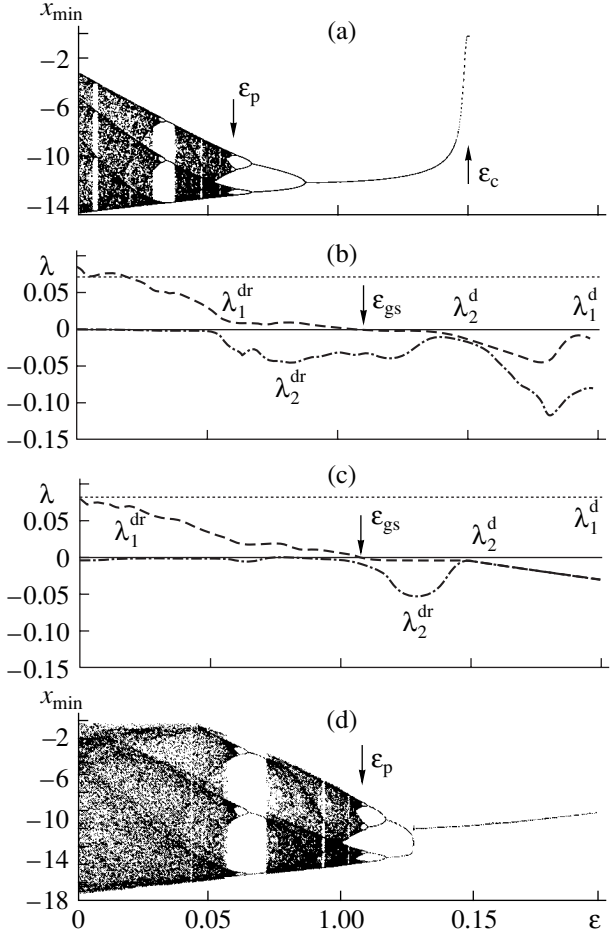

**Fig. 1.** (a) Bifurcation diagram for the modified Ressler system vs. control parameter $\varepsilon$, (b, c) $\varepsilon$ dependences of the Lyapunov exponent spectra for the Ressler system weakly ($\omega_d = 0.99$) and strongly ($\omega_d = 1.30$) offset in parameter $\omega$ (the conditional Lyapunov exponents are shown by the dashed, $\lambda_1^{dr}$, and dash-and-dot, $\lambda_2^{dr}$, lines), and (d) bifurcation diagram for the nonautonomous modified Ressler system subjected to an external harmonic perturbation. The value of $\varepsilon$ at which the periodic regime sets in, $\varepsilon_p$, is shown by the arrow.

Generalized synchronization occurs in the set of unidirectionally coupled stochastic oscillators only when the periodic regime in modified system (7) becomes sufficiently stable (the corresponding value of the coupling parameter, $\varepsilon = \varepsilon_{gs} \approx 0.11$, is indicated by the arrow in Fig. 1b). At such a value of the coupling parameter, a cycle of period one is realized in modified Ressler system (7). Note that, when the periodic regime is observed in the modified system (and, accordingly, generalized synchronization is established in the set of unidirectionally coupled stochastic oscillators), major conditional Lyapunov exponent $\lambda_1^{dr}$ is weakly negative. When the coupling parameter reaches some critical value, $\varepsilon_c \approx 0.15$ (the arrow in Fig. 1a), the modified Ressler system passes into the steady state and the major conditional Lyapunov exponent starts decreasing rapidly (cf. Figs. 1a and 1b).

Note that such an approach to treating the effect of generalized synchronization, which is based on considering the dynamics of the modified system, also gives an explanation for the fact [25] that the generalized synchronization threshold is virtually independent of the offsets of the control parameters of coupled stochastic oscillators. From our consideration, it follows that the stability of the periodic regime (which is necessary for generalized synchronization to take place) depends primarily on the intrinsic properties of modified system $\mathbf{x}_m(t)$. The value of $\varepsilon_{gs}$ that meets the onset of generalized synchronization depends on the offset of control parameter $\omega$ insignificantly (cf. the values of $\varepsilon_{gs}$ for $\omega_d = 0.99$ in Fig. 1b and $\omega_d = 1.3$ in Fig. 1c). As was mentioned above, this statement agrees well with the results reported in [25].

Let us now see why the value of control parameter $\varepsilon_{gs}$ at which generalized synchronization sets in coincides with none of the bifurcation points in the modified system (cf. Figs. 1a and 1b). The reason for such a noncoincidence is the perturbing effect of the driving oscillator. It was already noted that an external harmonic perturbation may generate chaotic oscillations in systems with periodic dynamics. In this case, the external perturbation shifts the bifurcation points of the nonautonomous modified system toward greater $\varepsilon$ compared with the dynamics of the autonomous modified system and the onset of generalized synchronization in the former does not coincide with the bifurcation points of the latter.

To clarify this conclusion, consider the behavior of the modified system subjected to an external harmonic perturbation,

$$\dot{x}_m = -\omega_{dr} y_m - z_m - \varepsilon x_m + A\cos(\Omega t),$$
$$\dot{y}_m = \omega_{dr} x_m + a y_m, \qquad (8)$$
$$\dot{z}_m = p + z_m(x_m - c),$$

where the values of the control parameters, $A = 1.32$ and $\Omega = 1.0$, are taken so as to simulate the dynamics of the driving oscillator. The bifurcation diagram for the nonautonomous modified system is shown in Fig. 1d. It is distinctly seen that all the bifurcation points of nonautonomous modified Ressler system (8) are shifted relative to the bifurcation points of autonomous modi-fied system (7) toward larger $\varepsilon$ (cf. Figs. 1a and 1d) in agreement with the aforesaid. The onset of generalized synchronization for the two Ressler systems is indicated by the arrow (Fig. 1b): generalized synchronization is established at such values of $\varepsilon$ when periodic oscillations arise in the nonautonomous modified system under the action of the harmonic perturbation.

The same effect causes the generalized synchronization regime in dynamic systems with discrete time (of mapping). It is known, for example, that generalized synchronization occurs in a set of unidirectionally coupled logistic maps at $\varepsilon \geq \varepsilon_{gs} \approx 0.32$ [20],

$$x_{n+1} = f(x_n),$$
$$y_{n+1} = f(y_n) + \varepsilon(f(x_n) - f(y_n)), \quad (9)$$

where $f(x) = 4x(1-x)$. Following the above approach, consider the modified system

$$z_{n+1} = (1-\varepsilon)f(z_n) = az_n(1-z_n), \quad (10)$$

where $a = 4(1-\varepsilon)$. It is clear that the value $\varepsilon_{gs} \approx 0.32$, at which generalized synchronization arises, corresponds to $a \approx 2.72$ for modified system (10). At such a value of the control parameter in logistic mapping, the stationary stable point $x^0 = (a-1)/a$ is an attractor.

## 3. GENERALIZED SYNCHRONIZATION IN CHAOTIC SYSTEMS WITH NONDISSIPATIVE COUPLING

Consider now mechanisms of generalized synchronization in the case of dissimilar unidirectionally coupled dynamic systems, including those with nondissipative coupling. Several systems of such a type are known [20, 22]. Clearly, in the case of dissipative coupling, the dissimilarity between the driving and driven systems is of minor significance and the aforesaid (see Sect. 2) remains valid in this case too. If, however, coupling is nondissipative, the modified system approach fails. An example of such a system is unidirectionally coupled Lorentz and Ressler oscillators [20]. As a driving system, a stochastic Ressler oscillator,

$$\dot{x}_d = -\alpha(y_d + z_d),$$
$$\dot{y}_d = \alpha(x_d + ay_d), \quad (11)$$
$$\dot{z}_d = \alpha(p + z_d(x_d - c)),$$

with parameters $\alpha = 6$, $a = 0.2$, $p = 0.2$, and $c = 5.7$ is used; as a driven system, a Lorentz oscillator,

$$\dot{x}_{dr} = \sigma(y_{dr} - x_{dr}),$$
$$\dot{y}_{dr} = rx_{dr} - y_{dr} - x_{dr}z_{dr} + \varepsilon y_d, \quad (12)$$
$$\dot{z}_{dr} = -bz_{dr} + x_{dr}y_{dr},$$

with parameters $\sigma = 10$, $r = 28$, and $b = 8/3$. Parameter $\alpha$ in Eq. (11) serves to change the characteristic scale of oscillations in the Ressler system. Coupling parameter $\varepsilon$ at which generalized synchronization is established between oscillators (11) and (12) was estimated in [20], $\varepsilon_{gs} \approx 6.66$. Then, oscillation amplitude $y_d$, the variable of the driving Ressler oscillator, roughly equals 10 dimensionless units and oscillation amplitude $y_{dr}$, the variable of the driven Lorentz oscillator, is about 20 units. Obviously, in this case, the amount of external signal $\varepsilon_{gs}y_d$ acting on Lorentz system (12) exceeds the amplitude of intrinsic oscillations in this system roughly by three times. Such a situation is depicted in Fig. 2, where time realization $y_{dr}(t)$ of the driven Lorentz system in the autonomous regime and external perturbation $\varepsilon_{gs}y_d(t)$ are shown. It is seen that the external force shifts the representative point in the phase space of the driven system toward domains with strong dissipation, as a result of which the intrinsic chaotic dynamics of the system becomes suppressed and the generalized synchronization conditions set in.

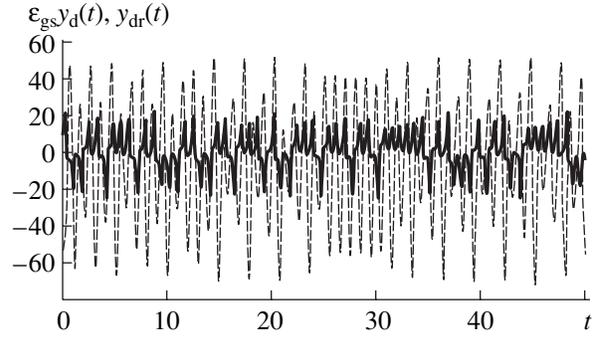

**Fig. 2.** Time realization $y_{dr}(t)$ corresponding to the autonomous dynamics of Lorentz system (12) (continuous line) and external perturbation $\varepsilon_{gs}y_d(t)$ (dashed line) introduced into the driven system at the point of onset of generalized synchronization. The amplitude of the external perturbation exceeds the amplitude of autonomous oscillations of the driven system by several times.

Thus, two similar mechanisms underlie the generalized synchronization regime, which are based on suppressing intrinsic chaotic oscillations by means of dissipation. This is accomplished either by introducing an additional dissipative term or by shifting the representative point of the system toward domains with strong dissipation in the phase space.

## 4. GENERALIZED SYNCHRONIZATION AND NOISE-INDUCED SYNCHRONIZATION

In a number of chaotic systems, noise-induced synchronization has been found [26–28]. Let two independent identical chaotic systems $\mathbf{u}(t)$ and $\mathbf{v}(t)$ subject to different initial conditions $\mathbf{u}(t_0)$ and $\mathbf{v}(t_0)$ fall into the range of the same chaotic attractor. Random signal $\xi(t)$ applied to them may "synchronize" the systems with each other; i.e., after the transient has been complete, the systems start to behave identically, $\mathbf{u}(t) \equiv \mathbf{v}(t)$. Therein lies the essence of noise-induced synchroniza-tion. Here again, as in the generalized synchronization regime, the synchronous dynamics of two systems shearing a noise source can be established only if the conditional Lyapunov exponents are negative [29, 30].

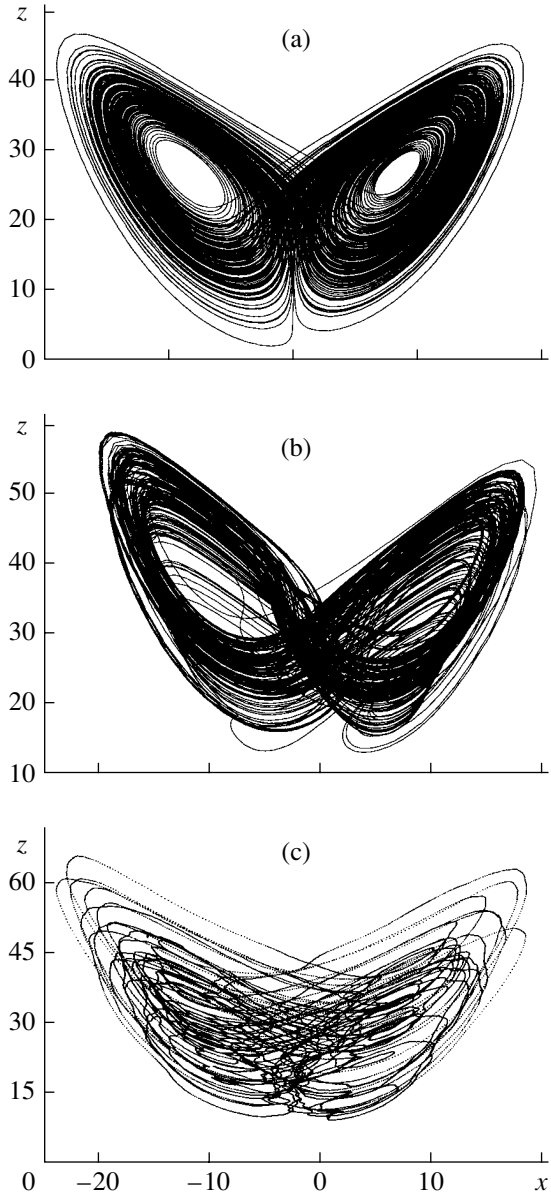

**Fig. 3.** Phase portrait of (a) autonomous Lorentz system (12) and (b) Lorentz system (12) locked in generalized synchronism with Ressler system (11) ($\varepsilon = 20$). (c) Noise-induced synchronization of Lorentz system (13) (system (12) subjected to random delta-correlated Gaussian process $\langle\xi(t)\xi(t')\rangle = \delta(t-t')$ with amplitude $\varepsilon = 40$) [28].

As was shown in [26–28], two similar mechanisms may be responsible for noise-induced synchronization.

(1) Random signal $\xi(t)$ applied to identical chaotic systems has a nonzero mean. In this case, the systems behave in a regular manner [31–33], merely "following" random external perturbation $\xi(t)$.

(2) An intense external signal (which may have a zero mean) transfers the representative point into the domains in the phase space that feature a high compression of the phase flux and the point stays in these domains for a long time. As a result, adjacent trajectories converge on average [28, 34, 35]. In both cases, the phase flux compression plays a decisive role and the conditional Lyapunov exponents are negative.

Although generalized chaotic synchronization and noise-induced synchronization are, as a rule, viewed as different phenomena, they occur by the same mechanism: suppression of intrinsic chaotic oscillations by dissipation (i.e., by introducing noise with a nonzero mean in the case of noise-induced synchronization and by introducing a dissipative term or by transferring the representative point of the system into high-dissipation domains of the phase space in the case of generalized synchronization).

Figure 3 shows the phase portraits for the (a) autonomous chaotic Lorentz system and (b, c) the same system subjected to an external perturbation. The generalized synchronization of oscillators (11) and (12) at $\varepsilon = 20$ is shown in Fig. 3b; the noise-induced synchronization, in Fig. 3c [28]. Here, the Lorentz system is described by the equations

$$\dot{x}_{dr} = \sigma(y_{dr} - x_{dr}),$$
$$\dot{y}_{dr} = rx_{dr} - y_{dr} - x_{dr}z_{dr} + \epsilon\xi(t), \quad (13)$$
$$\dot{z}_{dr} = -bz_{dr} + x_{dr}y_{dr}$$

(with the same values of the control parameters as in (12)) and random delta-correlated Gaussian process $\langle\xi(t)\xi(t')\rangle = \delta(t-t')$ with amplitude $\epsilon = 40$ is taken as an external perturbation. It is distinctly seen that the structure of the attractor is identical in both cases, since the synchronous behavior is caused by the same reason (for details concerning noise-induced synchronization, see [28]).

Thus, one can suppose that noise-induced synchronization may also be established in the case of dissipative coupling (like that used in relationships (6) and (9)) if a random process is applied instead of variable $x_d(t)$ of the driving system. By way of example, consider driven logistic mapping (9) for the case when the variation of quantity $x_n$ with discrete time, instead of being specified by evolution operator (9), is a random process $\xi_n$ with probability density $p(\xi)$. Then, the dynamics of the driven system is given by

$$y_{n+1} = f(y_n) + \varepsilon(f(\xi_n) - f(y_n)). \quad (14)$$

Let us show that, although quantity $\xi(t)$ is random, this random process and the dynamic system may lock in synchronism, with the synchronization being similar to generalized chaotic synchronization.

To diagnose generalized synchronization between random process $\xi_n$ and dynamic system $y_n$, we will invoke the modified system approach described above. Figure 4b shows the behavior of the driven and auxiliary systems ($y_n$ and $v_n$, respectively) for control param-eter $a = 3.75$ and random quantity $\xi$ obeying the normal law,

$$p(\xi) = \frac{1}{\sqrt{2\pi}\sigma}\exp\left(-\frac{(\xi - \xi_0)^2}{2\sigma^2}\right), \quad (15)$$

where $\xi_0 = 0.5$ and $\sigma = 0.11$.

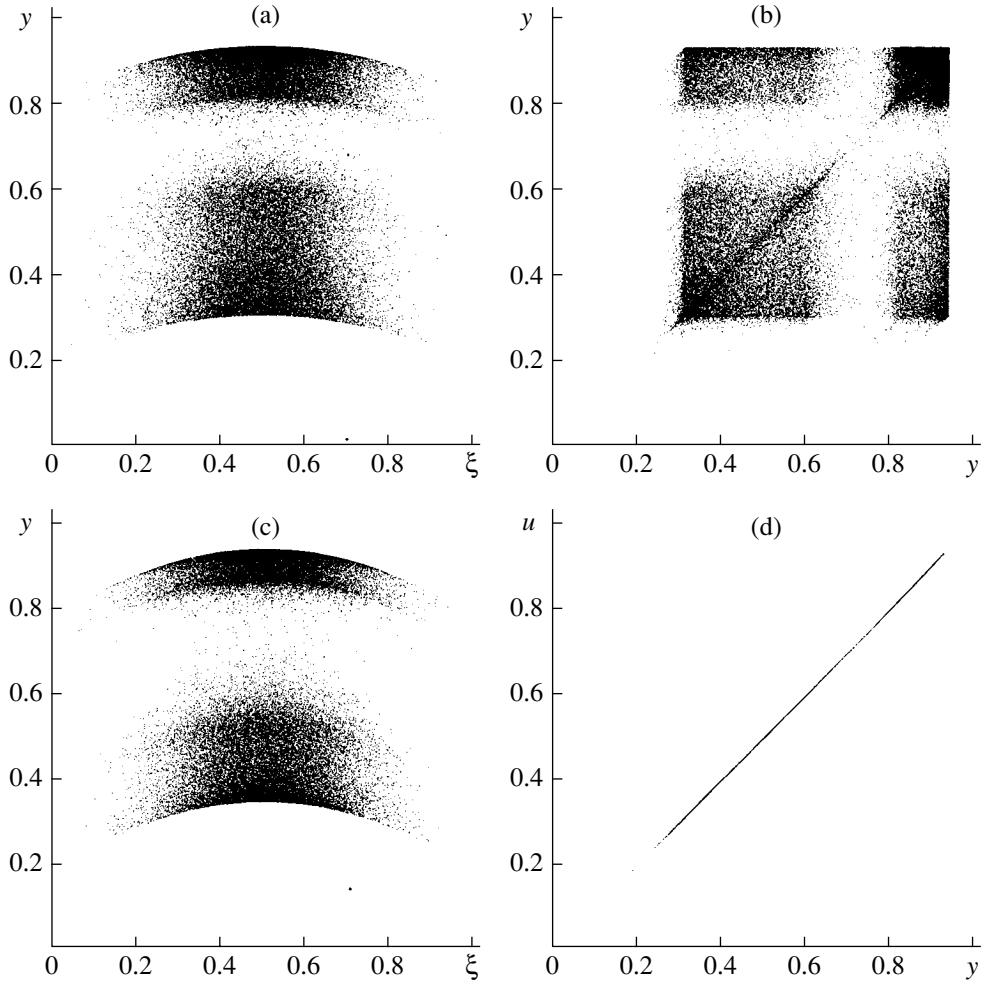

**Fig. 4.** Planes $(\xi_n, y_n)$ and $(y_n, v_n)$ of logistic mapping (14) for coupling parameter $\varepsilon =$ (a, b) 0.125 and (c, d) 0.175. At $\varepsilon = 0.175$, the driven, $y_n$, and auxiliary, $v_n$, systems behave in the same manner, $y_n = v_n$, indicating the presence of functional dependence $y_n = F(\xi_n)$ and, hence, establishment of the synchronous regime.

It is seen that, when the coupling parameter is small ($\varepsilon = 0.125$), the driven and auxiliary systems take on different values at the same instant of discrete time (the points characterizing the states of the system are distributed over the plane $(y, v)$); hence, there is no functional dependence between random process $\xi_n$ and a state of dynamic system $y_n$. When the coupling parameter increases to 0.175, the situation changes radically (Fig. 4d). The points corresponding to the states of the systems fall on the diagonal $y = v$, indicating the functional dependence $y_n = F(\xi_n)$; hence, the synchronous behavior of the systems takes place. Note that the func-tional dependence $\mathbf{F}[\cdot]$ is this case has a complicated fractal structure (Fig. 4c), which means that the synchronization is weak. Analyzing the plane $(\xi, y)$ alone, one cannot establish a functional correspondence (cf. Figs. 4a and 4c).

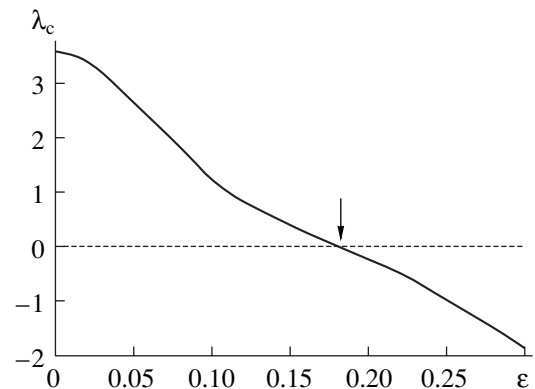

**Fig. 5.** Conditional Lyapunov exponent $\lambda_c$ of system (14) vs. coupling parameter $\varepsilon$. The distribution of Random quantity $\xi_n$ obeys normal law (15). The arrow marks the onset of synchronization.

The presence of synchronization is also corroborated by the dependence of conditional Lyapunov exponent $\lambda_c$ on coupling parameter $\varepsilon$ (Fig. 5). When the coupling parameter is small, conditional Lyapunov exponent $\lambda_c$ is positive, which means that a functional dependence between random quantity $\xi_n$ and a state of dynamic system $y_n$ is absent. As the coupling parameter grows, the conditional Lyapunov exponent becomes negative; hence, functional dependence $y_n = \mathbf{F}[\xi_n]$ and, accordingly, the synchronous regime are established.

## CONCLUSIONS

Thus, we found the reasons for generalized chaotic synchronization. To elucidate the underlying mechanisms, the method of auxiliary system was used. It was shown that the driven system behaves in accordance with the behavior of the modified system (with additional dissipation) subjected to a random external perturbation. An increase in the coupling parameter is equivalent to an increase in dissipation and the perturbation amplitude. The generalized synchronization threshold in this case depends on the tradeoff between the suppression of the intrinsic chaotic dynamics in the driven system and the excitation of chaotic oscillations in it under the action of an external signal (coming from the driving system). Similar processes associated with the suppression of the intrinsic dynamics also take place in establishing generalized synchronization in the case of nondissipative coupling, when an external perturbation shifts the representative point toward high-compression domains in the phase space.

It can also be concluded that generalized chaotic synchronization and noise-induced synchronization, while traditionally viewed as phenomena of different nature, take place for one reason: suppression of intrinsic chaotic oscillations by introducing dissipation.

## ACKNOWLEDGMENTS

This work was supported by the Ministry of Education and Science of the Russian Federation (project no. 333), Russian Foundation for Basic Research (project nos. 05-02-16273 and 05-02-16286), "Project in Support of Leading Scientific Schools" (project no. NSh-1250.2003.02), and CRDF (grant no. REC-006).

The authors are indebted to the Dinastiya Foundation and the International Center of Fundamental Physics (Moscow).